\documentclass[9pt,twocolumn,twoside]{opticajnl-arxiv}
\journal{ao} % Choose journal (ao,jocn,josaa,josab,ol,optica,pr)

\setboolean{shortarticle}{false}
\usepackage{lineno}
\usepackage{mathtools}
\usepackage{siunitx}
\newcommand*\diff{\mathop{}\!\mathrm{d}}
\newcommand*{\expv}[1]{\ensuremath{\langle #1 \rangle}}

\DeclareMathOperator*{\Var}{Var}
\title{Radiant fluence from ray tracing in optical multipass systems}

\author[1,2*]{Mirosław Marszałek}
\author[1]{Lukas Affolter}
\author[1,2]{Oguzhan Kara}
\author[1,2]{Klaus Kirch}
\author[1]{Karsten Schuhmann}
\author[1]{Manuel Zeyen}
\author[1,2]{Aldo Antognini}

\affil[1]{Institute for Particle Physics and Astrophysics, ETH Zurich, 8093 Zurich, Switzerland}
\affil[2]{Laboratory for Particle Physics, Paul Scherrer Institute, 5232 Villigen PSI, Switzerland}

\affil[*]{Corresponding author: miroslaw.marszalek@psi.ch}

\begin{abstract}
    In applications of optical multipass cells 
    in photochemical reactors and laser excitation of weak transitions,
    estimation of the radiation dose in a volume of interest 
    allows to assess the performance and optimize the design of the cell.
    We adopt radiant fluence as the figure of merit
    and employ the radiative transfer equation
    to derive analytical expressions 
    for average radiant fluence in a given volume of interest.
    These are used to establish practical approximations 
    and a Monte Carlo ray tracing model of the spatial distribution of fluence.
    Ray tracing is performed with Zemax OpticsStudio 18.9.
\end{abstract}

\setboolean{displaycopyright}{true}

\begin{document}

\maketitle

\section{Introduction}

Optical multipass cells are an established tool 
in absorption spectroscopy~\cite{hodgkinsonOpticalGasSensing2012,fathyDirectAbsorptionPhotoacoustic2022},
allowing to improve the signal thanks to increased optical path length through a sample.
There are, however, applications where it is their ability to concentrate radiation energy
rather than to extend the path length that is interesting,
although both goals can be seen as equivalent to some extent.
Multipass arrangements have been used to enhance excitation probabilities
of weak transitions in molecules~\cite{loeschStericEffectsState1993,riedelSimpleEffectiveMultipass2008,leavittIsomerSpecificIRIR2012}
and light exotic atoms~\cite{vogelsangMultipassLaserCavity2014,
pohlSizeProton2010,
antogniniProtonStructureMeasurement2013,
pohlLaserSpectroscopyMuonic2016,
krauthMeasuringAparticleCharge2021}.
Notably, all of the three upcoming measurements of the ground-state
hyperfine splitting in muonic hydrogen will feature a multipass cell~\cite{kandaPrecisionLaserSpectroscopy2018,pizzolottoFAMUExperimentMuonic2020,amaroLaserExcitation1shyperfine2022}.
In photochemical reactors~\cite{blatchleyPhotochemicalReactorsTheory2022},
they are utilized to increase the exposure of a chemical or biological sample 
to light~\cite{jensenInactivationAirborneViruses1964,
navntoftEffectivenessSolarDisinfection2008,
liImpactReflectionFluence2012,
thatcherImpactSurfaceReflection2021}, typically UV radiation.
The aim of this work is to provide means to model the performance
and guide the design of such optical systems.

Radiation dose in photochemistry is usually quantified with \emph{radiant fluence},
defined at a point in space as the radiant energy \(\diff Q\) 
incident directly on the surface of an infinitesimally small sphere surrounding that point
divided by the cross-sectional area \(\diff \sigma\) of the sphere~\cite{nicodemusSelfstudyManualOptical1978},
that is, \(F = \diff Q\!\,/\!\diff \sigma\), 
expressed as energy per unit area (\(\unit{\joule/\centi\metre\squared}\)).
The excitation rate of an atomic or molecular transition is determined by
the \emph{radiant energy density} \(w\)~\cite{loudonQuantumTheoryLight2000},
given by \(F = c \int w \diff t\),
where \(c\) is the speed of light in the medium 
and the integral runs over the time of illumination.
Under stationary illumination conditions, 
\emph{radiant fluence rate} \(\diff F\!/\!\diff t = cw\) may be used instead.

Ray tracing has been proposed~\cite{lauUltravioletIrradianceMeasurement2012,
ahmedRayTracingFluence2018,lombiniDesignOpticalCavity2021,houSpatialAnalysisImpact2021}
as an alternative to other commonly used approaches 
to compute fluence or fluence rate distributions~\cite{liuEvaluationAlternativeFluence2004,sunReviewFluenceDetermination2022}
thanks to its ability to handle large numbers of reflections in a multipass system
and ease of implementation with commercial ray tracing software.
However, 
precise explanations of how ray tracing data can be interpreted
to compute fluence or fluence rate seem to be difficult to find in literature,
and some subtle details are often overlooked.
In this work, we derive the relevant formalism directly from
the radiative transfer equation (RTE)~\cite{howellIntroductionRadiativeTransfer2021,modestRadiativeTransferEquation2022},
an integro-differential equation 
governing the transfer of electromagnetic radiation
including effects of absorption, emission 
and scattering on surfaces and in participating media.
Values of average fluence and spatial distributions of fluence
are extracted from Monte Carlo ray tracing~\cite{mahanMonteCarloRayTrace2019}
data of the optical system under consideration.
Furthermore, we show how approximate analytical expressions 
for average fluence in a volume of interest can be constructed
leveraging symmetries of the system.
Usage of the methods proposed here is illustrated on two examples of multipass cells,
with non-sequential ray tracing performed with Zemax OpticsStudio.

The central concept in our formulation is the path length of a ray through a volume,
referred to as the \emph{chord length}.
When computing spatial distributions of fluence from ray tracing,
one is concerned with chord lengths through small volume elements,
which may not be directly accessible in the ray tracing software used.
If spherical volume elements are used,
the chord length can be approximated by its average value~\cite{lauUltravioletIrradianceMeasurement2012,ahmedRayTracingFluence2018},
independently of the angular distribution of illumination.
Rectangular voxel grids are another common choice~\cite{nataleRaytracing3DDust2017,
    blancoFluxTracerRayTracer2019,
    lombiniDesignOpticalCavity2021,
    houSpatialAnalysisImpact2021},
simplifying ray tracing considerably at the cost of directional dependence.
In this case the chord length is often taken as the width of a voxel.
Authors in~\cite{lombiniDesignOpticalCavity2021}, for instance, 
suggest that the voxel width is a good approximation 
for collimated rays incident perpendicularly on a face of a voxel
or isotropic illumination.
We believe this assertion holds only in the former case
and show that the error in the latter may be as large as \qty{50}{\percent}.
Other approaches not relying on the chord length include
the method of cubic illumination~\cite{houSpatialAnalysisImpact2021},
which is known to exhibit large variation dependent on illumination conditions
of up to \qty{42}{\percent}~\cite{cuttleCubicIllumination1997,xiaDeterminingScalarIlluminance2023}.

The need to correctly account for the chord length through a voxel
has been long recognized in X-ray dosimetry,
where photon fluence in Monte Carlo methods is often computed
with the so-called \emph{track-length estimator}~\cite{williamsonMonteCarloEvaluation1987,baldacciTrackLengthEstimator2015},
used in several popular Monte Carlo codes~\cite{demarcoAnalysisMCNPCrosssections2002,chibaniMCPISubminuteMonte2005,mittoneEfficientNumericalTool2013}.
The concept originates in neutron transport theory~\cite{weinbergPhysicalTheoryNeutron1958,spanierTwoPairsFamilies1966}.
Similarities between the RTE and the linear Boltzmann equation,
describing transport of neutral particles~\cite{caseLinearTransportTheory1967},
allow us to borrow several ideas developed in this field.
We employ those to define the chord length distribution in a volume
and approximate average chord lengths under isotropic illumination.

The nomenclature does not seem to be well established.
While authors in the field of radiometry tend to refer to \(F\)
as \emph{radiant fluence}~\cite{nicodemusSelfstudyManualOptical1978,mccluneyIntroductionRadiometryPhotometry2014},
in photobiology it is more commonly known as \emph{energy fluence}~\cite{rupertDosimetricConceptsPhotobiology1974,bjoernPhotobiologyScienceLife2008}.
\emph{Spherical radiant exposure} has been used as well~\cite{braslavskyGlossaryTermsUsed2007,slineyRadiometricQuantitiesUnits2007},
highlighting its relationship with the more familiar quantity of \emph{radiant exposure},
defined as radiant energy incident on a given surface per unit projected area.
In this work, we refer to \(F\) as simply \emph{fluence}.

\section{Theoretical background}\label{sec:theory}
We assume a typical setting for applications 
in excitation of weak atomic or molecular transitions
and UV illumination in photochemical reactors:
monochromatic light and no light sources 
or re-emission within the illuminated sample.
Spatial and directional distribution of radiation is then described
by \emph{radiance} \(L\)~\cite{mccluneyIntroductionRadiometryPhotometry2014},
defined for a point \(\mathbf{r}\) and direction \(\mathbf{\Omega}\), 
\(\lVert \mathbf{\Omega} \rVert = 1 \), as
\begin{equation}
    L(t,\mathbf{r},\mathbf{\Omega}) = \frac{\diff Q /\!\diff t}{\diff \mathnormal{\Sigma}_\perp \diff \mathnormal{\Omega}}
    \quad\left[\unit[per-mode=fraction]{\watt\per\metre\squared\per\steradian}\right],
\end{equation}
where \(\diff Q\!\,/\!\diff t\) is the radiant energy per unit time 
flowing in the direction \(\mathbf{\Omega}\) 
through a surface element \(\diff \mathnormal{\Sigma}_\perp\) 
perpendicular to \(\mathbf{\Omega}\) 
into a solid angle \(\diff \mathnormal{\Omega}\).
For phenomena sufficiently slow for \(\partial L/c\partial t \approx 0\) to be valid,
which is the case in most engineering applications besides ultrafast physics,
distribution of radiance is governed by
the steady-state radiative transfer equation (RTE)~\cite{howellThermalRadiationHeat2021}
\begin{multline}\label{eq:RTE-L}
    \boldsymbol{\nabla} L(t,\mathbf{r},\mathbf{\Omega})\cdot\mathbf{\Omega} = \\
-(\mu_\text{a} + \mu_\text{s})L(t,\mathbf{r},\mathbf{\Omega}) 
+ \mu_\text{s}\int_{4\pi}L(t,\mathbf{r},\mathbf{\Omega}')\varphi(\mathbf{\Omega}',\mathbf{\Omega})\diff \mathnormal{\Omega}'.
\end{multline}
Here \(\mu_\text{a}\) and \(\mu_\text{s}\) are the absorption and scattering coefficients, respectively, 
and \(\varphi(\mathbf{\Omega}',\mathbf{\Omega})\) is the scattering phase function,
defined as the probability of scattering from direction \(\mathbf{\Omega}'\)
into direction \(\mathbf{\Omega}\).
The LHS of \eqref{eq:RTE-L} is the directional derivative of \(L\) 
along the direction \(\mathbf{\Omega}\),
the first term of the RHS stands for losses due to absorption and out-scattering,
and the second term represents the gain due to 
in-scattering of light from other directions.

The relevant quantities of interest can be derived from radiance \(L\)~\cite{nicodemusSelfstudyManualOptical1978}.
Radiant fluence rate is obtained from
\begin{equation}\label{eq:dF-L}
    \frac{\diff F(t,\mathbf{r})}{\diff t} = \int_{4\pi} L(t,\mathbf{r},\mathbf{\Omega}) \diff \mathnormal{\Omega}
    \quad\left[\unit[per-mode=fraction]{\watt\per\metre\squared}\right],
\end{equation}
whereas radiant energy density is given by
\begin{equation}
    w(t,\mathbf{r}) = \frac{1}{c}\frac{\diff F(t,\mathbf{r})}{\diff t}
    \quad\left[\unit[per-mode=fraction]{\joule\per\metre\cubed}\right].
\end{equation}
Radiant fluence \(F\) is related to radiance \(L\) through
\begin{equation}\label{eq:F-L}
    F(t,\mathbf{r}) = \int_0^t\int_{4\pi}
    L(t',\mathbf{r},\mathbf{\Omega}) \diff \mathnormal{\Omega} \diff t',
\end{equation}
where \(t=0\) is chosen as the beginning of illumination.
By introducing the \emph{radiance dose} \(G\), defined as
\begin{equation}
    G(t,\mathbf{r},\mathbf{\Omega}) = \int_0^t L(t',\mathbf{r},\mathbf{\Omega}) \diff t'
    \quad\left[\unit[per-mode=fraction]{\joule\per\metre\squared\per\steradian}\right],
\end{equation}
\eqref{eq:F-L} becomes
\begin{equation}
    F(t,\mathbf{r}) = \int_{4\pi} G(t,\mathbf{r},\mathbf{\Omega}) \diff \mathnormal{\Omega}
    \quad\left[\unit[per-mode=fraction]{\joule\per\metre\squared}\right].
\end{equation}
Under the assumption of relatively slow phenomena, 
the RTE for radiance dose \(G\) is formally identical to \eqref{eq:RTE-L}
and can be obtained by integrating it over time, yielding
\begin{multline}\label{eq:RTE-G}
        \boldsymbol{\nabla} G(t,\mathbf{r},\mathbf{\Omega})\cdot\mathbf{\Omega} = \\
    -(\mu_\text{a} + \mu_\text{s})G(t,\mathbf{r},\mathbf{\Omega}) 
    + \mu_\text{s}\int_{4\pi}G(t,\mathbf{r},\mathbf{\Omega}')\varphi(\mathbf{\Omega}',\mathbf{\Omega})\diff \mathnormal{\Omega}'.
\end{multline}
Therefore, all results for fluence \(F\) presented in the following sections 
apply to fluence rate \(\diff F\!/\!\diff t\) and radiant energy density \(w\)
by a simple analogy.

\subsection{Average fluence in a volume}
Radiant fluence may be found by integrating \(G\) over the full solid angle,
which in turn can be obtained by solving \eqref{eq:RTE-G}.
This, however, is a hard problem in a general setting.
Consider instead the average fluence in a given volume of interest \(V\)
\begin{equation}
    \expv{F} = \frac{1}{V}\int_V\int_{4\pi} G(\mathbf{r},\mathbf{\Omega}) \diff \mathnormal{\Omega} \diff V.
\end{equation}
We may carry out the integration over \(\mathnormal{\Omega}\) and \(V\) 
on both sides of \eqref{eq:RTE-G}.
Since the phase function is normalized such that
\begin{equation}
    \int_{4\pi} \varphi(\mathbf{\Omega}',\mathbf{\Omega}) \diff \mathnormal{\Omega} = 1,
\end{equation}
the RHS of \eqref{eq:RTE-G} yields
\begin{equation}
    -(\mu_\text{a} + \mu_\text{s})V\expv{F} + \mu_\text{s} V\expv{F} = -\mu_\text{a} V\expv{F}.
\end{equation}
On the LHS we employ the product rule for divergence
\begin{equation}
    \boldsymbol{\nabla}\cdot(G(\mathbf{r},\mathbf{\Omega})\mathbf{\Omega})
    = \boldsymbol{\nabla} G(\mathbf{r},\mathbf{\Omega})\cdot\mathbf{\Omega} 
    + G(\mathbf{r},\mathbf{\Omega})\underbrace{\boldsymbol{\nabla}\cdot\mathbf{\Omega}}_{=0}
\end{equation}
and use the divergence theorem, resulting in
\begin{equation}
    \int_V\int_{4\pi} \boldsymbol{\nabla} G(\mathbf{r},\mathbf{\Omega})\cdot\mathbf{\Omega} \diff \mathnormal{\Omega} \diff V 
    = -\oint_{\partial V}\int_{4\pi} G(\mathbf{r},\mathbf{\Omega}) \mathbf{\Omega}\cdot\mathbf{n} \diff \mathnormal{\Omega} \diff \mathnormal{\Sigma},        
\end{equation}
where \(\mathbf{n}\) is the inward normal on the volume boundary \(\partial V\).
Splitting the integral over solid angles into the outward and inward part,
the above can be interpreted as
\begin{multline}
    -\Bigg(\oint_{\partial V}\int_{\mathbf{\Omega}\cdot\mathbf{n}\geq 0} G(\mathbf{r},\mathbf{\Omega}) \mathbf{\Omega}\cdot\mathbf{n} \diff \mathnormal{\Omega} \diff \mathnormal{\Sigma} \\
        -\oint_{\partial V}\int_{\mathbf{\Omega}\cdot\mathbf{n}<0} G(\mathbf{r},\mathbf{\Omega}) \mathbf{\Omega}\cdot(-\mathbf{n}) \diff \mathnormal{\Omega} \diff \mathnormal{\Sigma}\Bigg) =\\
    -(Q_\text{in} - Q_\text{out}) = -Q_\text{abs},
\end{multline}
where \(Q_\text{in}\) and \(Q_\text{out}\) 
are the radiant energies entering and leaving the volume, respectively,
and thus \(Q_\text{abs}\) is the energy absorbed within the volume.
Therefore, reconstituting both sides yields
\begin{equation}\label{eq:Favg-Qabs}
    \expv{F} = \frac{Q_\text{abs}}{\mu_\text{a} V},
\end{equation}
which does not depend on scattering properties of the medium.

\begin{figure}[!ht]
    \centering
    \includegraphics{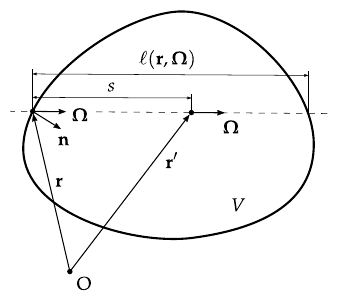}
    \caption{
        Integration of radiance over a volume of interest \(V\) can be replaced
        by integration over its boundary \(\partial V\).
        To do so, we relate radiance at a point \(\mathbf{r}'\in V\)
        with its value at the point \(\mathbf{r}=\mathbf{r}'-s\mathbf{\Omega}\in\partial V\).
        The resulting expression involves the chord length \(\ell(\mathbf{r},\mathbf{\Omega})\),
        which is the distance from \(\mathbf{r}\) to the opposing part of \(\partial V\)
        along the direction \(\mathbf{\Omega}\).
    }\label{fig:chord}
\end{figure}

A solution in another form can be derived from the integrated RTE~\cite{howellThermalRadiationHeat2021}.
For simplicity, the discussion is limited here to non-scattering media,
i.e., for \(\mu_\text{s} = 0\).
In this regime, the integrated RTE for radiance dose connects values of \(G\)
along a line of sight through
\begin{equation}\label{eq:RTE-int}
    G(\mathbf{r}',\mathbf{\Omega}) = G(\mathbf{r},\mathbf{\Omega})\exp(-\mu_\text{a} s),
\end{equation}
where \(\mathbf{r}'=\mathbf{r}+s\mathbf{\Omega}\)
and \(s = \lvert\mathbf{r}'-\mathbf{r}\rvert\) is the path length along the line of sight.
In particular, for \(\mathbf{r}\in\partial V\) and \(\mathbf{r}'\in V\)
this equation connects values of \(G\) between the boundary of \(V\)
and its interior, as illustrated in Fig.~\ref{fig:chord}.
The average fluence in \(V\) is then found by integrating over all solid angles
and all \(\mathbf{r}'\) within \(V\), that is,
\begin{multline}
    \expv{F} = \frac{1}{V} \int_V\int_{4\pi} G(\mathbf{r}',\mathbf{\Omega}) \diff \mathnormal{\Omega} \diff V' = \\
    \frac{1}{V} \int_V\int_{4\pi} G(\mathbf{r},\mathbf{\Omega})\exp(-\mu_\text{a} s) \diff \mathnormal{\Omega} \diff V'.
\end{multline}
The integral over \(\mathbf{\Omega}\) can be interpreted 
as integration over \(\mathbf{r}\in\partial V\) 
with \(\diff \mathnormal{\Omega} = \cos\theta \diff \mathnormal{\Sigma} / s^2\),
where \(\cos\theta = \mathbf{\Omega}\cdot \mathbf{n}\).
For a given point \(\mathbf{r}\in\partial V\),
the integral over \(\mathbf{r}'\in V\) can be then decomposed into integration
over solid angles in a hemisphere and over \(s\) within boundaries of \(V\)
with \(dV' = s^2 \diff s \diff \mathnormal{\Omega}\).
The resulting expression is
\begin{equation}
    \expv{F} = \frac{1}{V} \oint_{\partial V}\int_{2\pi}\int_0^{\ell(\mathbf{r},\mathbf{\Omega})}
     G(\mathbf{r},\mathbf{\Omega})\exp(-\mu_\text{a} s) \cos\theta \diff s \diff \mathnormal{\Omega} \diff \mathnormal{\Sigma},
\end{equation}
where \(\ell(\mathbf{r},\mathbf{\Omega})\) is the chord length, i.e.,
the distance from \(\mathbf{r}\) along \(\mathbf{\Omega}\) 
to the opposing boundary of \(V\).
The integral over \(s\) can be carried out explicitly, leading to
\begin{equation}\label{eq:Favg-full}
    \expv{F} = \frac{1}{\mu_\text{a} V} \oint_{\partial V}\int_{2\pi}
     \Bigl(1 - \exp\bigl(-\mu_\text{a} \ell(\mathbf{r},\mathbf{\Omega})\bigr)\Bigr)G(\mathbf{r},\mathbf{\Omega}) \cos\theta \diff \mathnormal{\Omega} \diff \mathnormal{\Sigma}.
\end{equation}
For a weakly-absorbing medium, 
the series expansion of \(\exp(-\mu_\text{a}\ell) \approx 1 - \mu_\text{a}\ell\) yields
\begin{equation}\label{eq:Favg-approx}
    \expv{F} = \frac{1}{V} \oint_{\partial V}\int_{2\pi}
    \ell(\mathbf{r},\mathbf{\Omega}) G(\mathbf{r},\mathbf{\Omega}) \cos\theta \diff \mathnormal{\Omega} \diff \mathnormal{\Sigma}.
\end{equation}
This relationship connect the average fluence in a volume
to illumination of its boundary described by \(G(\mathbf{r},\mathbf{\Omega})\)
and its geometry via \(\ell(\mathbf{r},\mathbf{\Omega})\).

\subsection{Average chord length}
Note that with the total radiant energy entering the volume
\begin{equation}\label{eq:Q-def}
    Q \equiv Q_\text{in} = \oint_{\partial V}\int_{2\pi} G(\mathbf{r},\mathbf{\Omega}) \cos\theta \diff \mathnormal{\Omega} \diff \mathnormal{\Sigma},
\end{equation}
the function
\begin{equation}\label{eq:pdf}
    p(\mathbf{r},\mathbf{\Omega})
    = \frac{G(\mathbf{r},\mathbf{\Omega}) \cos\theta}{Q}
\end{equation}
may be understood as a probability density.
Eqs.~(\ref{eq:Favg-full},~\ref{eq:Favg-approx}) 
can be simplified by introducing the moments
\begin{align}
    \expv{e^{-\mu_\text{a}\ell}} &= \oint_{\partial V}\int_{2\pi}
    \exp\bigl(-\mu_\text{a} \ell(\mathbf{r},\mathbf{\Omega})\bigr) p(\mathbf{r},\mathbf{\Omega}) \diff \mathnormal{\Omega} \diff \mathnormal{\Sigma}, \label{eq:exp-chord-avg}\\
    \expv{\ell} &= \oint_{\partial V}\int_{2\pi}
    \ell(\mathbf{r},\mathbf{\Omega}) p(\mathbf{r},\mathbf{\Omega}) \diff \mathnormal{\Omega} \diff \mathnormal{\Sigma}, \label{eq:chord-avg}
\end{align}
leading to 
\begin{equation}\label{eq:Favg-exp-chord}
    \expv{F} = \frac{Q\left(1 - \expv{e^{-\mu_\text{a}\ell}}\right)}{\mu_\text{a} V}
\end{equation}
for an absorbing medium, and
\begin{equation}\label{eq:Favg-chord}
    \expv{F} = \frac{Q\expv{\ell}}{V}
\end{equation}
in the weak absorption regime.
\(\expv{\ell}\) can be interpreted as the average chord length
weighted by the radiance dose on the boundary of the volume of interest.

Similar concepts have been used in neutron transport theory~\cite{weinbergPhysicalTheoryNeutron1958,caseIntroductionTheoryNeutron1953}
to determine escape probabilities in nuclear reactors.
To advance this analogy even further,
let us follow the idea of Dirac 
(as cited in~\cite{caseIntroductionTheoryNeutron1953})
and define a chord length distribution function \(\phi(\ell)\)
such that the probability of finding a chord of length between \(\ell\) and \(\ell+\diff\ell\) is
\begin{equation}\label{eq:chord-dist}
    \phi(\ell)\diff\ell 
    = \oint_{\partial V}\int_{\ell}p(\mathbf{r},\mathbf{\Omega}) \diff \mathnormal{\Omega} \diff \mathnormal{\Sigma},
\end{equation}
where the second integral is carried out only over directions 
for which the chord has a length between \(\ell\) and \(\ell+\diff\ell\).
The moments in Eqs.~(\ref{eq:exp-chord-avg},~\ref{eq:chord-avg}) then become
\begin{align}
    \expv{e^{-\mu_\text{a}\ell}} &= \int_0^\infty \exp\bigl(-\mu_\text{a}\ell\bigr) \phi(\ell) \diff\ell, \\
    \expv{\ell} &= \int_0^\infty \ell \phi(\ell) \diff\ell.
\end{align}

The case of isotropic neutron flux,
analogous to isotropic illumination \(G(\mathbf{r},\mathbf{\Omega}) = G(\mathbf{r})\),
attracted a particularly high attention~\cite{dekruijfAverageChordLength2003,sanchezUseAverageChord2004}.
Chord length distributions and their moments 
have been derived for a variety of geometrical shapes~\cite{stoyanStochasticGeometryIts2013,gilleParticleParticleSystems2013}
in this regime.
Notably, the average chord length of a body in three dimensions
is given by the well-known formula attributed to Cauchy,
\begin{equation}\label{eq:cauchy-3d}
    \expv{\ell} = \frac{4V}{\mathnormal{\Sigma}},
\end{equation}
where \(V\) is the volume of the body and \(\mathnormal{\Sigma}\) is its surface area.
The analogous formula in two dimensions, i.e., for a plane figure, is
\begin{equation}\label{eq:cauchy-2d}
    \expv{\ell} = \frac{\pi A}{C},
\end{equation}
where \(A\) is the area of the figure and \(C\) is its perimeter.
These results are not restricted to convex bodies~\cite{mazzoloPropertiesChordLength2003,vlasovExtensionDiracChord2011}.
Furthermore, as pointed out in~\cite{dekruijfAverageChordLength2003},
the ratio \(V/\expv{\ell} = \mathnormal{\Sigma}/4\)
is the projected area of the body under isotropic illumination.
In the general case, we may interpret it as the average projected area
weighted by the radiance dose \(G\).
Defining
\begin{equation}\label{eq:proj-avg}
    \expv{\sigma} \equiv V/\expv{\ell}
\end{equation}
allows to rewrite \eqref{eq:Favg-chord} as
\begin{equation}\label{eq:Favg-sigma}
    \expv{F} = \frac{Q}{\expv{\sigma}}.
\end{equation}
Note the similarity of this result to the definition of fluence \(F = \diff Q\!\,/\!\diff \sigma\).

While we have neglected scattering in the derivation presented in the previous section,
\eqref{eq:cauchy-3d} has been shown to remain valid
under isotropic illumination in scattering media~\cite{bardsleyAverageTransportPath1981,blancoInvariancePropertyDiffusive2003}.
Furthermore, note that from Eqs.~(\ref{eq:Favg-Qabs}) and~(\ref{eq:Favg-exp-chord},~\ref{eq:Favg-chord}),
\begin{equation}
    Q_\text{abs} = Q\left(1 - \expv{e^{-\mu_\text{a}\ell}}\right),
\end{equation}
which reproduces the Beer--Lambert law and reduces to 
\begin{equation}
    Q_\text{abs} = Q\mu_\text{a}\expv{\ell}
\end{equation}
in a weakly-absorbing medium.
Since \eqref{eq:Favg-Qabs} is derived from the full RTE,
and thus \(Q_\text{abs}\) is independent of scattering properties of the medium,
we may extend the above expressions to include scattering
by interpreting \(\expv{\ell}\) as the average path length
through the volume of interest including scattering,
or \(\ell(\mathbf{r},\mathbf{\Omega})\) in Eqs.~(\ref{eq:Favg-full},~\ref{eq:Favg-approx})
as the average path length of a light ray scattering through the volume 
starting at \((\mathbf{r},\mathbf{\Omega})\).
A rigorous derivation of this result involves a Liouville--Neumann series
solution of the full integrated RTE~\cite{stuartMultipleScatteringNeutrons1957,xiao-linNewFastMonte2021}
and is outside the scope of this work.
While this interpretation is perhaps unfeasible for analytical calculation,
it lends itself very well to Monte Carlo ray tracing.

\section{Methods}\label{sec:methods}
In the present section, we describe three approaches to compute the average fluence
or fluence distribution in a given volume of interest:
\begin{enumerate}
    \itemindent4EM
    \item[Method \ref{sec:methods-direct}:] Monte Carlo ray tracing relying on
    appropriate sampling of the light source
    and MC estimates of Eqs.~(\ref{eq:Favg-full},~\ref{eq:Favg-approx}).
    \item[Method \ref{sec:methods-avg}:] Using~Eqs.~(\ref{eq:Favg-chord},~\ref{eq:Favg-sigma})
    directly when the incident radiant energy and average chord length can be inferred by some means,
    for instance from Cauchy formulas~(\ref{eq:cauchy-3d},~\ref{eq:cauchy-2d}) 
    for isotropic illumination.
    \item[Method \ref{sec:methods-spatial}:] Via Monte Carlo ray tracing on a voxel grid,
    where one of the above methods is applied to each voxel,
    producing a spatial distribution of fluence within the volume of interest.
\end{enumerate}
These methods are applied to two examples of multipass systems 
in Sec.~\ref{sec:results}.

Note that none of these methods can account for the wave nature of light,
essentially assuming incoherent illumination.
In particular, when the purpose of calculation is 
to approximate the excitation probability,
interference within the volume of interest may locally increase values of fluence
beyond the weak excitation regime and introduce saturation effects.
Nevertheless, thanks to conservation of energy,
computation of the average fluence should remain valid
as long as dimensions of the volume of interest are considerably larger
than the length scale of interference features of the illumination pattern.

\subsection{Monte Carlo ray tracing}\label{sec:methods-direct}
We employ commercial software, Zemax OpticsStudio 18.9,
to carry out Monte-Carlo ray tracing, 
relieving us from many implementation details.
Nevertheless, some crucial aspects are described below for completeness.
In the following,
a ray is defined as a path from the source to its termination point.
Rays are terminated when they leave the system 
or their energy drops below the predefined threshold, 
typically \num{1e-3} of their initial energy.
A ray segment is a section of a ray between two successive points of interaction,
be it on real or virtual surfaces or at points of scattering events.
A ray thus consists of numerous connected segments.

The computation begins with sampling of the light source 
according to its radiance dose distribution \(G(\mathbf{r}_0,\mathbf{\Omega}_0)\),
where \(\mathbf{r}_0\) lie on the surface of the source.
In case of a surface source,
\(N_0\) rays are initialized with equal energies \((Q_0)_i=Q_0/N_0\),
where \(Q_0\) is the initial energy of the beam.
Their coordinates \((\mathbf{r}_0,\mathbf{\Omega}_0) \) are sampled from the probability density
\begin{equation}
    p(\mathbf{r}_0,\mathbf{\Omega}_0)
    = \frac{G(\mathbf{r}_0,\mathbf{\Omega}_0) \cos\theta_0}{Q_0},
\end{equation}
where \(\cos\theta_0 = \mathbf{\Omega}_0\cdot\mathbf{n}_0\) 
for the outward surface normal \(\mathbf{n}_0\) at \(\mathbf{r}_0\)
and angles \(\theta_0\in[0,\pi/2]\).
A volumetric source may be defined in a similar manner.
OpticsStudio includes numerous models of light sources
and allows to implement user-defined source distributions.

After initialization, rays are propagated according to laws of geometrical optics.
There are several approaches to model realistic effects 
of absorption and scattering in ray tracing~\cite{farmerComparisonMonteCarlo1998,howellMonteCarloMethod2021,modestMonteCarloMethod2022}.
OpticsStudio utilizes the so-called \emph{pathlength method},
which treats absorption deterministically.
A ray segment of length \(x\) passing through 
a medium of an absorption coefficient \(\mu_\text{a}\)
has its energy attenuated by a factor of \(\exp(-\mu_\text{a} x)\).
Absorption due to reflections on partially-reflecting surfaces is implemented similarly.
This method effectively reduces variance of the result,
as randomness of photon absorption events is marginalized out.

Volumetric scattering is modeled by sampling the free path \(x\) 
from the probability density \(p(x) = 1 - \exp(-\mu_\text{s} x)\), 
after which a scattering event occurs.
On a scattering event, either in a volume or on a surface,
a new ray segment is initialized 
in a direction chosen according to the given probability distribution.
Additionally, when scattering on a surface,
the ray may be split into several rays instead, which is known as \emph{ray splitting}.
However, we opt for \emph{ray redirection} instead
since in a Monte-Carlo simulation of typical size
ray splitting may result in excessive numbers of rays.
The program includes numerous scattering models for surface and volume scattering,
and allows to implement arbitrary scattering phase functions.

Radiation transfer from the source to the volume of interest
determines the radiance dose \(G(\mathbf{r},\mathbf{\Omega})\) at the volume boundary, 
which may be associated with the probability density \(p(\mathbf{r},\mathbf{\Omega})\) in \eqref{eq:pdf}.
The ray tracing procedure results in a sampling of this density
with a set of ray intersections \((\mathbf{r}_i,\mathbf{\Omega}_i)\).
Since absorption is treated deterministically,
ray energies at intersections \(Q_i\) act as weighting factors,
that is, each intersection may be associated with \(N_i\propto Q_i\) samples,
and since the total radiant energy entering the volume is \(Q = \sum Q_i\), 
we have \(N_i/N = Q_i/Q\) for \(N = \sum N_i\).
Furthermore, each intersection is associated with a chord length \(\ell_i\),
which is the total length of all segments comprising the section of the ray
from the intersection to the point of exit from the volume.
Note that each ray may cross the volume of interest multiple times,
thus generating multiple intersections.

Monte Carlo estimates of the relevant quantities may be then computed
by replacing integrals involving \(p(\mathbf{r},\mathbf{\Omega})\)
with average values of these quantities
since \((\mathbf{r}_i,\mathbf{\Omega}_i)\) are sampled from \(p(\mathbf{r},\mathbf{\Omega})\).
For instance, the average chord length in \eqref{eq:chord-avg} can be expressed as
\begin{multline}
    \expv{\ell} = \oint_{\partial V}\int_{2\pi}
    \ell(\mathbf{r},\mathbf{\Omega}) p(\mathbf{r},\mathbf{\Omega}) \diff \mathnormal{\Omega} \diff \mathnormal{\Sigma} \\
    = \frac{1}{N} \sum_i \ell_i N_i = \frac{1}{Q} \sum_i \ell_i Q_i,
\end{multline}
which is simply the energy-weighted mean value.
Similarly, \eqref{eq:exp-chord-avg} becomes
\begin{align}\label{eq:exp-chord-avg-rt}
    \expv{e^{-\mu_\text{a}\ell}} &= \frac{1}{Q} \sum_i \exp(-\mu_\text{a}\ell_i) Q_i.
\end{align}
The average fluence in Eqs.~(\ref{eq:Favg-full},~\ref{eq:Favg-approx})
is then estimated by
\begin{equation}\label{eq:Favg-chord-exp-rt}
    \expv{F} = \frac{1}{\mu_\text{a} V} \sum_i  \Bigl(1 - \exp\bigl(-\mu_\text{a} \ell_i\bigr)\Bigr) Q_i
    = \frac{1}{\mu_\text{a} V} \sum_i (Q_\text{abs})_i
\end{equation}
in an absorbing medium,
where \((Q_\text{abs})_i\) is the energy absorbed from the \(i\)-th chord, and
\begin{equation}\label{eq:Favg-chord-rt}
    \expv{F} = \frac{1}{V} \sum_i \ell_i Q_i
\end{equation}
in the weakly absorbing case.
Equations~(\ref{eq:Favg-chord-exp-rt},~\ref{eq:Favg-chord-rt})
are the foundation of our ray trace model of radiant fluence.
Finally, the chord length distribution \(\phi(\ell)\)
can be approximated through the empirical probability of finding a chord
of length between \(\ell-\Delta\ell\) and \(\ell+\Delta\ell\) as~\cite{dekkingModernIntroductionProbability2005}
\begin{equation}\label{eq:empirical-chord-dist}
    \phi(\ell) = \frac{1}{2\Delta\ell Q} 
    \sum\limits_{\ell_i > \ell-\Delta\ell 
                 \atop \ell_i \leq\ell+\Delta\ell}\mkern-12mu\ell_i Q_i.
\end{equation}

To quantify the uncertainty of the Monte Carlo estimate of average fluence,
we divide sampled source rays into \(M\) groups of \(N_0/M\) rays each
and compute the average fluence \(\expv{F_m}\) for each group,
where \(m = 1,\ldots,M\).
From Eqs.~(\ref{eq:Favg-chord-exp-rt},~\ref{eq:Favg-chord-rt}) it is clear
that this amounts to dividing the sums into \(M\) partitions, hence
\begin{equation}
    \expv{F} = \sum_{m=1}^M \expv{F_m}.
\end{equation}
Variance of \(\expv{F}\) may be then estimated 
from the sample variance of \(\expv{F_m}\) as
\begin{equation}\label{eq:mc-variance}
    \Var\left[\expv{F}\right] = M\Var\left[\expv{F_m}\right]
    = \frac{1}{M(M-1)}\sum_{m=1}^M \Bigl(M\expv{F_m} - \expv{F} \Bigr)^2.
\end{equation}

While the resulting formulas are straightforward, 
the information about chord lengths \(\ell_i\) is not easily accessible in OpticsStudio.
It is necessary to export a ray database after a ray trace,
which may be a relatively large file,
and analyze it using external tools,
making the process cumbersome and computationally expensive.
However, the program does report the total energy \(Q_\text{abs} = \sum (Q_\text{abs})_i\)
absorbed in a volume containing an absorbing medium,
hence one may use \eqref{eq:Favg-Qabs} directly instead.
In a non-absorbing medium, an artificial absorption coefficient \(\mu_\text{a}\)
of a value sufficiently small to ensure that other sources of energy loss
dominate can be introduced.
In other words, we choose \(\mu_\text{a}\) such that \(\mu_\text{a}\ell_\text{max}\ll\expv{\eta}\),
where \(\ell_\text{max}\) is the maximum chord length through the volume of interest
and \(\eta\) is the average relative energy loss per pass,
for instance, due to limited reflectivity of surfaces
or absorption in other parts of the system.

\subsection{Approximate average chord length}\label{sec:methods-avg}
In this method the average fluence is calculated from \eqref{eq:Favg-chord} 
by estimating \(Q\) and \(\expv{\ell}\) from properties of the multipass system.

In a multipass cell,
the total energy might be established from the reflectivity of mirrors \(R\)
and the number of passes in the cell \(n\).
Let \(Q_0\) be the initial energy of a light beam injected into the cell.
The beam energy on each pass follows the geometric sequence
\begin{equation}
    Q_0,\ Q_0R,\ Q_0R^2,\ \ldots,\ Q_0R^{n-1}.
\end{equation}
Assuming that all the light passes through the volume of interest,
the total energy entering the volume is given by the sum of this sequence
\begin{equation}
    Q = Q_0\,\frac{1 - R^n}{1 - R}.
\end{equation}

\begin{figure}[!ht]
    \centering
    \includegraphics{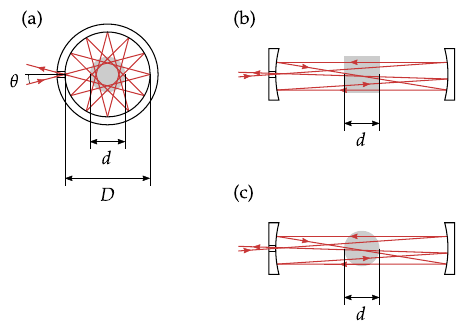}
    \caption{
        Example configurations of multipass cells and volumes of interest
        (marked with grey regions),
        where simple approximations of the average chord length can be used.
        In a circular cell and a disk-shaped volume (a) and
        a Herriott cell with a volume shaped as a box or cylinder (b),
        the chord length of the beam is nearly constant on each pass.
        In a Herriott cell with a disk-shaped volume of interest (c),
        we may approximate the average chord length 
        by that of a uniform distribution.
    }\label{fig:methods-avg-chord}
\end{figure}

The average chord length could be inferred, 
for instance, from symmetry of the system.
Two extreme cases can be recognized,
which are illustrated in examples displayed in Fig.~\ref{fig:methods-avg-chord}.

First, in certain geometries \(\ell\) is almost constant,
hence it may substituted for the average chord length.
This can happen, for instance, 
in a circular cell of diameter \(D\)
and a disk-shaped volume of interest of diameter \(d\)
shown in Fig.~\ref{fig:methods-avg-chord}a,
since the angle of incidence \(\theta\) is conserved in such a cell.
The average chord length may be then approximated by the chord of a circle
\begin{equation}\label{eq:chord-circ}
    \expv{\ell} = \sqrt{d^2 - D^2\sin^2\theta},
\end{equation}
which is \(\expv{\ell}\approx d\) for small \(\theta\).
Similarly, in a Herriott cell and a volume of interest
of a shape of a box or cylinder of length \(d\)
pictured in Fig.~\ref{fig:methods-avg-chord}b,
the average chord length can be approximated by \(\expv{\ell}\approx d\),
as here the beam travels almost parallel to the axis
and has nearly the same path length through the volume on every pass.

Second, in some situations the distribution of chord lengths is nearly uniform.
One may then use the Cauchy formula \eqref{eq:cauchy-3d}
or, when propagation takes place mostly in a plane,
its two-dimensional version given by \eqref{eq:cauchy-2d}.
In the example pictured in Fig.~\ref{fig:methods-avg-chord}c,
a Herriott cell is used with a disk-shaped volume of interest of diameter \(d\).
Due to symmetry of the circle and 
since the path length probes all values between \(0\) and \(d\),
we may presume that the average chord length will be close 
to that of isotropic illumination in two dimensions,
hence from \eqref{eq:cauchy-2d}
\begin{equation}\label{eq:chord-herriott}
    \expv{\ell} \approx \frac{\pi\cdot\pi d^2/4}{\pi d} = \frac{\pi d}{4}.
\end{equation}

\subsection{Ray tracing on a voxel grid}\label{sec:methods-spatial}
The two methods described so far can be applied to individual volume elements
of a voxel grid or other spatial sub-divisions of the volume of interest.
If the volume elements are sufficiently small,
values of their average fluence may be assigned to points in space,
thus creating a spatial map of fluence.
In other words, fluence at a point \(\mathbf{r}_k\) can be approximated as
\(F(\mathbf{r}_k) \approx \expv{F_k}\), 
where \(\expv{F_k}\) is given by Eqs.~(\ref{eq:Favg-chord-exp-rt},~\ref{eq:Favg-chord-rt})
for the volume \(V_k\) of the \(k\)-th element centered at \(\mathbf{r}_k\).

Method~\ref{sec:methods-direct} requires access to chord lengths
or energy \((Q_\text{abs})_k\) absorbed in each volume element.
In OpticsStudio, voxel grids are implemented as the \emph{Volume Detector Object},
which does not expose chord lengths but reports the total absorbed energy
in each voxel.
We therefore opted for the absorbed energy approach,
employing an artificial absorption coefficient 
when a non-absorbing medium was considered.

When only the total energy incident on each voxel is available,
method~\ref{sec:methods-avg} can be employed 
with a suitable approximation of the average chord length.
For sufficiently small volume elements, 
we may assume that the spatial ray density is uniform across the volume,
that is, \(G(\mathbf{r},\mathbf{\Omega}) = G(\mathbf{\Omega})\),
leaving us with the angular distribution of radiation.
Spherical volume elements are a convenient choice
since their projected area is independent of direction \(\mathbf{\Omega}\),
which allows to approximate the average chord length by that of isotropic illumination.
For a sphere of radius \(\rho\) \eqref{eq:cauchy-3d} yields \(\expv{\ell} = 4\rho/3\).
Note that this result is equivalent to the approach 
used by~\cite{lauUltravioletIrradianceMeasurement2012,ahmedRayTracingFluence2018},
who calculated fluence by dividing the energy incident 
on the spherical volume element by its cross-sectional area,
as the average projected area from \eqref{eq:proj-avg} is in this case
\(\expv{\sigma} = V\expv{\ell} = \pi\rho^2\).

Another common choice is a rectangular voxel grid, 
which simplifies ray tracing considerably,
although at the cost of dependence on the angular distribution of radiation.
Two extreme cases can be considered.
For collimated light incident perpendicularly on faces of voxels of width \(a\),
the chord length is well approximated with \(\expv{\ell} = a\).
The other extreme is isotropic illumination,
in which case the 3D Cauchy formula~(\ref{eq:cauchy-3d}) yields \(\expv{\ell} = 2a/3\).
Note that using \(\expv{\ell} = a\) for isotropic illumination
would lead to an overestimation of average fluence by \qty{50}{\percent}.
In an intermediate case of 2D isotropic illumination,
which may be realized when light propagation is confined mostly to a plane,
the 2D Cauchy formula~(\ref{eq:cauchy-2d}) gives \(\expv{\ell} = \pi a/4\).
We consider such an example in the following section.

\section{Results and Discussion}\label{sec:results}
To illustrate the methods of calculation of average fluence 
described in the previous section,
we apply them to two optical multipass systems:
a circular cell modeled after~\cite{tuzsonCompactMultipassOptical2013}
and a Herriott cell~\cite{herriottOffAxisPathsSpherical1964}.
In these examples, neither absorption nor scattering effects are considered.

\subsection{Models of the cells}
\begin{figure}[!ht]
    \centering
    \includegraphics{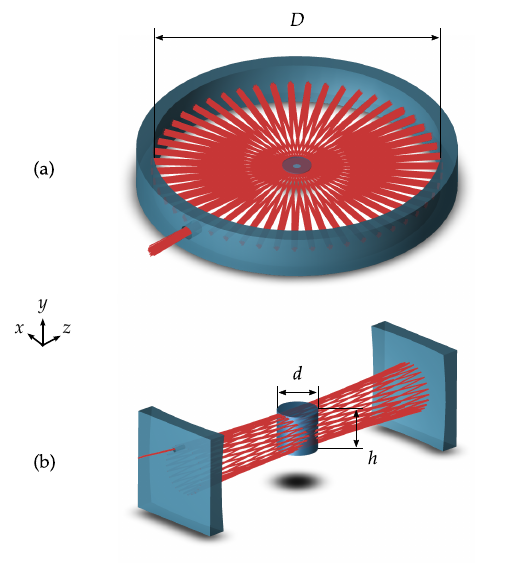}
    \caption{
        Models of the two multipass cells used as examples in this work:
        (a) circular cell and (b) Herriot cell.
    }\label{fig:cells-raytrace}
\end{figure}

Models of the cells were implemented in OpticsStudio using the non-sequential mode,
which allows to trace large ensembles of rays over multiple reflections.
Fig.~\ref{fig:cells-raytrace} shows renderings of the models.
The circular cell has a diameter of \(D = \qty{100}{\mm}\),
and the light is injected at an angle of \(\theta = \qty{1.8}{\degree}\)
chosen for 51 passes.
The Herriott cell consists of two spherical mirrors configured for 52 passes.
Mirror reflectivity in both cases is \(R = \qty{99.2}{\percent}\).
A gray solid in the center of each cell represents the volume of interest, 
which is a cylinder of a diameter \(d = \qty{10}{\mm}\) in both cases.
The height of the cylinder is \(h = \qty{1}{\mm}\) in the circular cell
and \(h = \qty{10}{\mm}\) in the Herriott cell.

The laser beam has a wavelength of \(\lambda = \qty{633}{\nm}\)
and is modeled as a Gaussian density of rays on phase space,
which follows the same propagation laws as a true Gaussian beam
in the paraxial regime~\cite{torreLinearRayWave2005}.
The beam waist is located in the center of the cell in both cases
and is \qty{10}{\um} and \qty{200}{\um} in the circular and Herriot cell,
respectively.
Each ray trace consisted of \(N=\num{1e4}\) rays of a total energy of \(Q_0=\qty{1}{\milli\joule}\)
distributed evenly among them.

\subsection{Results}
Values of average fluence \(\expv{F}\) obtained with each method described in Sec.~\ref{sec:methods}
are collected in Tab.~\ref{tab:results}.
Method \ref{sec:methods-spatial} is considered in two variants,
depending on whether Method \ref{sec:methods-direct} or \ref{sec:methods-avg}
was used to compute average fluence in a voxel.
Uncertainties were estimated from \eqref{eq:mc-variance} 
by dividing the calculation into \(M=100\) groups of \(N/M=100\) rays each.

\begin{table}[ht!]
    \centering
    \caption{Results of calculations of average fluence for both example optical systems
    carried out with the three methods described in Sec.~\ref{sec:methods}.}\label{tab:results}
    \begin{tabular}{
            lc
            S[table-format = 3.4(2)]
            S[print-implicit-plus = true, table-format = +2.2]
        }\hline
        Cell & Method                                               & {\(\expv{F}/\unit{\milli\joule\per\cm\squared}\)} & {Error\,/\,\unit{\percent}} \\\hline
    Circular & \ref{sec:methods-direct}                             & 509.43 +- 0.19        & {---}     \\
             & \ref{sec:methods-avg}                                & 508.93                & -0.10     \\
             & \ref{sec:methods-spatial}/\ref{sec:methods-direct}   & 509.72 +- 0.14        & 0.06      \\
             & \ref{sec:methods-spatial}/\ref{sec:methods-avg}      & 512.54 +- 0.27        & 0.61      \\[1EM]
    Herriott & \ref{sec:methods-direct}                             &  40.4997 +- 0.0086    & {---}     \\
             & \ref{sec:methods-avg}                                &  42.6778              & 5.38      \\
             & \ref{sec:methods-spatial}/\ref{sec:methods-direct}   &  40.562 +- 0.016      & 0.15      \\
             & \ref{sec:methods-spatial}/\ref{sec:methods-avg}      &  45.108 +- 0.013      & 11.38     \\\hline
    \end{tabular}
\end{table}

\begin{figure}[!ht]
    \centering
    \includegraphics{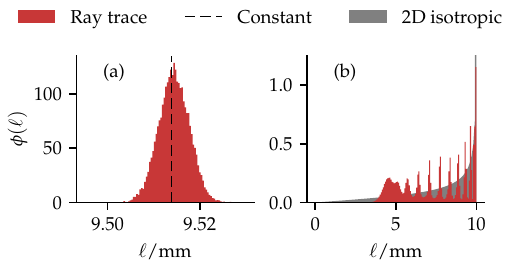}
    \caption{
       Chord length distributions obtained with Method~\ref{sec:methods-direct} (ray trace)
       for the circular (a) and Herriott (b) cells.
       Additionally, the constant chord length assumed for (a)
       and the 2D isotropic distribution for a circle
       assumed for (b) in Method~\ref{sec:methods-avg} are marked.
    }\label{fig:chord-dists}
\end{figure}

Method~\ref{sec:methods-avg} relies on estimation of \(\expv{\ell}\) from geometry of the problem.
For the circular cell, we take a constant chord length given by \eqref{eq:chord-circ},
whereas for the Herriott cell, we approximate it with uniform illumination in two dimensions
and calculate it from \eqref{eq:chord-herriott}.
To illustrate the accuracy of these assumptions,
empirical chord length distributions \(\phi(\ell)\) obtained with
Method~\ref{sec:methods-direct} using \eqref{eq:empirical-chord-dist}
are plotted in Fig.~\ref{fig:chord-dists}.
Distributions assumed in Method~\ref{sec:methods-avg}---a constant chord length
for the circular cell and a 2D isotropic distribution
for the Herriott cell---are marked for comparison.
The analytical form of the 2D isotropic distribution for a circle of a diameter \(d\)
is given by~\cite{colemanRandomPathsConvex1969}
\begin{equation}
    \phi(\ell) = \frac{\ell}{d\sqrt{d^2-\ell^2}},\quad 0<\ell<d.
\end{equation}

\begin{figure}[!ht]
    \centering
    \includegraphics{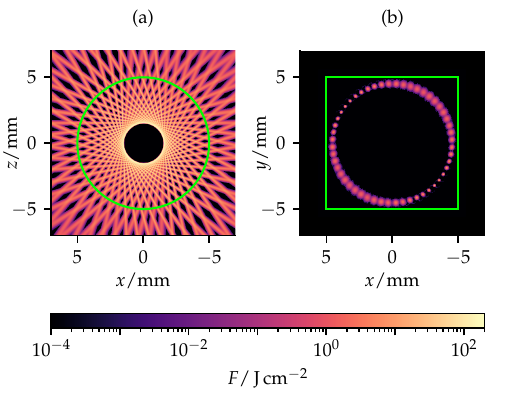}
    \caption{
        Sections through fluence distributions 
        for the circular (a) and Herriott (b) cells
        computed with Method~\ref{sec:methods-spatial}/\ref{sec:methods-direct}.
        Boundaries of the volume of interest are marked with the green line.
        Distributions are shown in logarithmic scale.
    }\label{fig:fluence-dists}
\end{figure}

For the spatial fluence distribution calculation with Method~\ref{sec:methods-spatial},
we used grids of voxels of widths of \(a = \qty{30}{\um}\) for the circular cell
and \(a = \qty{50}{\um}\) for the Herriott cell.
These were chosen to maximize the number of voxels in a volume detector
encompassing the volume of interest
within the limit of \num{1e7} voxels imposed by the software.
Note that this setting does not allow to resolve the beam at its waist
in the circular cell, hence the maximal obtained values of fluence are expected
to be somewhat underestimated.
In the variant~\ref{sec:methods-spatial}/\ref{sec:methods-direct},
we introduce an artificial absorption coefficient 
of \(\mu_\text{a} = \qty{1e-6}{\per\mm}\),
resulting in a relative energy loss of around \num{1e-5} 
per pass through the region of interest,
which is negligible in comparison with the loss of \num{8e-3}
per reflection on the mirror surface.
For the variant~\ref{sec:methods-spatial}/\ref{sec:methods-avg},
we assume 2D isotropic illumination in the circular cell,
as the propagation takes place mostly in a plane,
yielding an average chord length of a voxel of \(\expv{\ell} = \pi a/4\).
In the Herriot cell,
we take the chord length equal to the width of a voxel \(\expv{\ell} = a\).
Sections of spatial distributions
through the \(x\)-\(z\) plane in the circular cell
and the \(x\)-\(y\) plane in the Herriott cell
obtained with Method~\ref{sec:methods-spatial}/\ref{sec:methods-direct}
are displayed in Fig.~\ref{fig:fluence-dists}.
Boundaries of the volume of interest are marked with a green line.
The average fluence presented in Tab.~\ref{tab:results} was computed
as the average of all voxels contained within the volume of interest.

\subsection{Discussion}
In the presented analysis, Method~\ref{sec:methods-direct}
serves as a reference for the other two methods,
as it is a direct Monte-Carlo calculation of average fluence
derived from the formalism presented in Sec.~\ref{sec:theory}.

In the circular cell,
the distribution of chord lengths obtained with Method~\ref{sec:methods-direct}
(Fig.~\ref{fig:chord-dists}a) is strongly concentrated around the central value.
This allows to approximate the average chord length in Method~\ref{sec:methods-avg}
very well, resulting in a relatively accurate estimate of average fluence.
In the Herriott cell, 
the observed chord length distribution (Fig.~\ref{fig:chord-dists}b)
only remotely resembles the assumed 2D isotropic distribution.
Most of the discrepancy of around \qty{5}{\percent} stems from this error, 
while the rest is caused by inaccurate determination of the total radiant energy
due to some rays missing the volume of interest
or reflecting more than the designed number of passes.
The actual pattern in the distribution, 
and thus the difference between the two methods,
depends on configuration of the cell and may become larger.

While Method~\ref{sec:methods-spatial}/\ref{sec:methods-direct}
is in good agreement with the Monte-Carlo estimate,
Method~\ref{sec:methods-spatial}/\ref{sec:methods-avg} shows a discrepancy
of around \qty{11}{\percent} in the Herriott cell.
Clearly, the approximation of \(\expv{\ell}=a\) does not hold very well
since the beam is not exactly perpendicular to the voxel grid.
The error is much smaller in the circular cell;
however, thanks to the cylindrical symmetry of the system,
averaging fluence over the region of interest 
is expected to reduce the discrepancy.
Local values of fluence may thus remain relatively inaccurate.

Note that in the chosen examples 
the chord length through the volume of interest 
can be approximated in Method~\ref{sec:methods-avg} easily,
since light follows rather predictable paths.
For the same reason, 
tracing a full ray distribution of the beam 
in Method~\ref{sec:methods-direct} is not necessary,
as one could trace a single ray and assign all energy to it.
However, this method can be used with more advanced optical systems 
with complicated beam paths, including strongly diverging beams,
scattered light, etc., and with complex shapes of volumes of interest.
The examples shown here were chosen to be relatively simple
to illustrate the methods better.
We consider more elaborate situations in a following publication.

\section{Conclusion}
In this publication, we present several approaches to estimate
the average radiant fluence and fluence rate and their spatial distribution 
in optical multipass systems.
The methods are derived from the radiative transfer equation,
allowing to account for absorption and scattering effects
on surfaces and in participating media.
The basis of our formulation is a relationship of 
radiance dose on the closed boundary of a volume 
and the distribution of chord lengths through the volume
with the average fluence in its interior.

The main approach we propose is based on Monte Carlo ray tracing
of the light source radiance dose distribution,
which directly samples the radiance dose and chord length distribution 
of the volume of interest.
The second method attempts to estimate the average fluence
by inferring the average chord length from symmetry of the system.
The well-known Cauchy formulas for average chord lengths 
under isotropic illumination, originally used in neutron transport,
prove useful in this case.
The third approach applies either of the two methods to individual volume elements
of a voxel grid, producing a spatial map of fluence.
We implement ray tracing in commercial software, Zemax OpticsStudio 18.9.
The three methods are compared on two examples of optical multipass cells
with a laser source.

Our approaches can be utilized to predict the performance
of optical multipass systems for excitation of weak atomic transitions,
photochemical reactors, etc., 
or in other applications where the RTE or linear Boltzmann equation are used.

\begin{backmatter}
\bmsection{Funding} %Content in the funding section will be generated entirely from details submitted to Prism. Authors may add placeholder text in the manuscript to assess length, but any text added to this section in the manuscript will be replaced during production and will display official funder names along with any grant numbers provided. If additional details about a funder are required, they may be added to the Acknowledgments, even if this duplicates information in the funding section. See the example below in Acknowledgements.
The European Research Council (ERC) through CoG. \#725039; 
the Swiss National Science Foundation through the projects 
SNF 200021\_165854 and SNF 200020\_197052.

% \bmsection{Acknowledgments} 

\bmsection{Disclosures} 
The authors declare no conflicts of interest.

\bmsection{Data Availability Statement}
Data underlying the results presented in this paper 
are not publicly available at this time 
but may be obtained from the authors upon reasonable request.

% \bmsection{Supplemental document}
% See Supplement 1 for supporting content. 

\end{backmatter}

\bibliography{references}

% Full bibliography added automatically for Optics Letters submissions; the following line will simply be ignored if submitting to other journals.
% Note that this extra page will not count against page length
\bibliographyfullrefs{references}

\end{document}